\documentclass[runningheads]{llncs}
\usepackage{graphicx}
\usepackage{color}
\usepackage{dsfont}
\usepackage{amssymb}
\usepackage{algorithm}
\usepackage{algpseudocode}
\usepackage{multirow}

\begin{document}
\title{Self-Loop Uncertainty: A Novel Pseudo-Label for Semi-Supervised Medical Image Segmentation}
%
%
\author{Yuexiang Li \and Jiawei Chen \and Xinpeng Xie \and Kai Ma \and Yefeng Zheng}
\authorrunning{Y. Li et al.}
\institute{Tencent Jarvis Lab, Shenzhen, China\\
\email{vicyxli@tencent.com}
}
\authorrunning{}
%
%
\maketitle              
\begin{abstract}
    Witnessing the success of deep learning neural networks in natural image processing, an increasing number of studies have been proposed to develop deep-learning-based frameworks for medical image segmentation. However, since the pixel-wise annotation of medical images is laborious and expensive, the amount of annotated data is usually deficient to well-train a neural network. In this paper, we propose a semi-supervised approach to train neural networks with limited labeled data and a large quantity of unlabeled images for medical image segmentation. A novel pseudo-label (namely self-loop uncertainty), generated by recurrently optimizing the neural network with a self-supervised task, is adopted as the ground-truth for the unlabeled images to augment the training set and boost the segmentation accuracy. The proposed self-loop uncertainty can be seen as an approximation of the uncertainty estimation yielded by ensembling multiple models with a significant reduction of inference time. Experimental results on two publicly available datasets demonstrate the effectiveness of our semi-supervied approach.
    

\end{abstract}
\section{Introduction}
Deep neural networks often require large quantity of labeled images to achieve satisfactory performance. However, since annotating medical images requires experienced physicians to spend hours or days to investigate, which is laborious and expensive, the labeled medical images are often very deficient, especially for the tasks requiring pixel-wise annotations (e.g., segmentation). To tackle this problem, many researches \cite{BaiWJ2017,BortsovaG2019,SedaiS2019,YuL2019} have been proposed to improve the segmentation performance of deep neural networks through exploiting the information from unlabeled data. Using pseudo-labels of unlabeled data (generated automatically by a segmentation algorithm via uncertainty estimation) is one of the potential solutions, which has been extensively studied. The most popular approaches are: 1) softmax probability map \cite{BaiWJ2017}, 2) Monte Carlo (MC) dropout \cite{SedaiS2019,YuL2019}, and 3) uncertainty estimation via network ensemble \cite{LakshminarayananB2017}. Specifically, Bai et al. \cite{BaiWJ2017} proposed a semi-supervised approach for the cardiac magnetic resonance volume segmentation. The proposed approach first used a limited number of labeled data to train the neural network and then utilized the softmax probability maps predicted by the neural network as the pseudo-label for the unlabeled volumes to augment the training set. In a more recent study, Sedai et al. \cite{SedaiS2019} proposed an uncertainty guided semi-supervised learning framework for the segmentation of retinal layers in optical coherence tomopgraphy images. The pseudo-label for semi-supervised learning was generated using the Monte Carlo (MC) dropout \cite{GalY2016}, which can be viewed as an approximation of Bayesian uncertainty. Uncertainty estimation via model ensemble \cite{LakshminarayananB2017} is another form of approximation of Bayesian uncertainty, which separately trained $K$ networks and combined the softmax probability map of each network $k$ by averaging as the ensemble uncertainty (i.e., $\frac{1}{K} \sum_{k=1}^K p_{k}$, where $p$ is the probability map).

Due to the variety of existing uncertainty estimation methods, Jungo et al. \cite{JungoA2019} conducted experiments to evaluate the reliability and limitation of existing approaches and concluded several observations. Two of them cause our interests: 1) the widely-used MC-dropout-based approaches are heavily dependent on the influence of dropout on the segmentation performance; 2) the computational-expensive ensemble method yields the most reliable results and is typically a good choice if the resources allow. To this end, an efficient way to yield the reliable ensemble uncertainty is worthwhile to investigate.

In this paper, we propose a novel pseudo-label, namely self-loop uncertainty, for the semi-supervised medical image segmentation. The proposed self-loop uncertainty is generated by recurrently optimizing the encoder of a fully convolutional network (FCN) with a self-supervised sub-task (e.g., Jigsaw puzzles). The benefits of integrating self-supervised learning into our framework can be summarized in two folds: 1) the self-supervised learning sub-task encourages the neural network to deeply mine the information from raw data and benefits the image segmentation task; 2) the same network at different stages during the self-supervised sub-task optimization can be seen as different models, which leads our self-loop uncertainty to an approximation of ensemble uncertainty with much lower computational cost. We evaluate the proposed semi-supervised learning approach on two medical image segmentation tasks---nuclei segmentation and skin lesion segmentation. Experimental results show that our self-loop uncertainty can significantly improve the segmentation accuracy of the neural network, which outperforms the currently widely-used pseudo-label (e.g., softmax probability map and MC dropout).

\begin{figure}[!tb]
    \centering
    \includegraphics[width=0.95\textwidth]{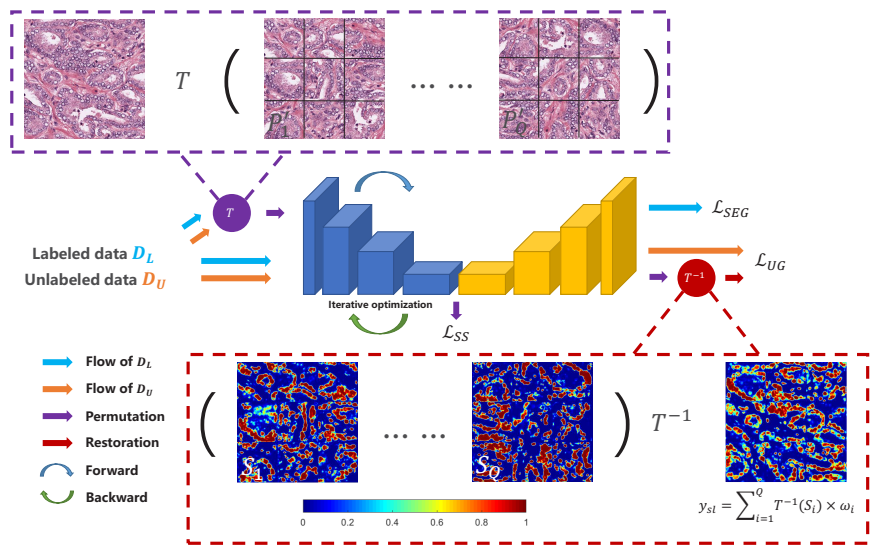}
    \caption{The pipeline of our semi-supervised segmentation framework. The proposed framework recurrently optimizes the encoder part of FCN by addressing the self-supervised learning task (i.e., supervised by $\mathcal{L}_{SS}$) to generate the pseudo-label for the unlabeled data. There are two losses, i.e., segmentation loss $\mathcal{L}_{SEG}$ and uncertainty-guided loss $\mathcal{L}_{UG}$, adopted in our framework to supervise the segmentation of labeled and unlabeled data. Our framework generates $Q$ permutations ($P_{1}^{'}$, ... $P_{Q}^{'}$) for an image (either labeled or unlabeled) and yields corresponding $Q$ segmentation predictions ($S_{1}$, ... $S_{Q}$) for the estimation of self-loop uncertainty $y_{sl}$ (as illustrated in Alg.~\ref{alg:ysl}).} \label{fig1:pipeline}
\end{figure}

\section{Method}
The proposed semi-supervised segmentation framework is illustrated in Fig.~\ref{fig1:pipeline}. The training set for our semi-supervised framework consists of labeled data $D_L$ and unlabeled data $D_U$. The proposed semi-supervised framework involves three losses (i.e., $\mathcal{L}_{SEG}$, $\mathcal{L}_{UG}$, and $\mathcal{L}_{SS}$) to supervise the network training with $D_L$ and $D_U$, respectively. The colored arrows in Fig.~\ref{fig1:pipeline} represent the information flows of $D_{U}$ (orange) and $D_{L}$ (cyan). For a batch containing images from $D_L$ and $D_U$, we calculate the supervised segmentation loss $\mathcal{L}_{SEG}$ (i.e., binary cross-entropy loss in our experiment) for labeled data with pixel-wise annotation to ensure the FCN has the segmentation capacity, self-supervised loss $\mathcal{L}_{SS}$ for both $D_L$ and $D_U$ to exploit rich information from raw data and generate the self-loop uncertainty, and uncertainty-guided loss $\mathcal{L}_{UG}$ for the unlabeled images to boost the segmentation performance of FCN with unlabeled data.




\subsection{Self-supervised Sub-task}
As aforementioned, the self-supervised loss $\mathcal{L}_{SS}$ aims to exploit rich information contained in raw data and generate the self-loop uncertainty. Various pretext tasks, such as rotation prediction \cite{gidaris2018image_rotations} and colorization \cite{larsson_colorization_2017}, can be adopted to achieve this goal. In this study, we use Jigsaw puzzles \cite{noroozi2016jigsaw_puzzles} consisting of translation and rotation transformations as the self-supervised sub-task to recurrently optimize the encoder of an FCN and yield the self-loop uncertainty.

Similar to the standard Jigsaw puzzles, we partition the image into several tiles, e.g., nine tiles for $3 \times 3$ Jigsaw puzzles. To formulate the Jigsaw puzzles sub-task, we permute the tiles using the approach proposed by \cite{noroozi2016jigsaw_puzzles}---a small subset $\mathds{P}^{'}$ of the large permutation pool, i.e., $\mathds{P}= (P_{1}, P_{2},...,P_{9!})$ is formed by selecting the $ K $ permutations with the largest Hamming distance between each other. In each training iteration, the input image is repeatedly disarranged (Q times in total where $Q \ll K$, $Q=10$ and $K=100$ in our experiments) by one randomly selected permutation from $\mathds{P}^{'}$.  Meanwhile, the encoder of FCN is recurrently updated to identify the selected permutation from the $K$ options for each disarranged image, which can be seen a classification task with $ K $ categories; therefore, we employ the cross-entropy loss as $\mathcal{L}_{SS}$ to supervise the sub-task.

\begin{algorithm}[!tb]
    \caption{Generation of self-loop uncertainty.}
    \label{alg:ysl}
    \begin{algorithmic}[1]
        \State \textbf{Input:}
        \State Network weights: $\theta_e$ of the encoder and $\theta_d$ of the decoder. \State Unlabeled data: $x \in D_U$.
        \State \textbf{Function:}
        \State $f(x; \theta)$ neural network forward function.
        \State $update(.)$ backpropagation to update the neural network weights.
        \State $T(.)$ permuted transformation of Jigsaw puzzles.
        \State $T^{-1}(.)$ inverse-permuted transformation.
        \State $\mathcal{L}_{SS}(p, g)$ calculation of the self-supervised loss with prediction $p$ and self-supervised signal $g$.
        \State \textbf{Procedure$^{\dagger}$:}
        \State $Q$ permutations are randomly selected from $\mathds{P}^{'}$: $\{P_1^{'},...,P_Q^{'} \in \mathds{P}^{'}\}$.
        \For {$i \in \{1,...,Q\}$}
        \State $p_i \leftarrow f(T_{P_i^{'}}(x); \theta_{e})$; $S_i \leftarrow f(T_{P_i^{'}}(x); \{\theta_{e},\theta_{d}\})$; 
        \State $l_i \leftarrow \mathcal{L}_{SS}(p_i, g_i)$; $\theta_{e}^i \leftarrow update(l_i)$; 
        \State $\theta_{e} \leftarrow \theta_{e}^i$. 
        \EndFor
        \State $y_{sl} = \sum_{i=1}^{Q} T^{-1}_{P_i^{'}}(S_i) \times norm(\omega_i)$, where $norm(.) = \frac {\omega_i} {\sum_{i=1}^Q\omega_i}$ and $\omega_i = 1-\frac {l_i} {\sum_{i=1}^Q l_i}$.
        \item[$\dagger$] $S$ is the segmentation prediction of FCN.
        \item[$\dagger$] $l$ is the calculated self-supervised loss.
        \State \textbf{Output:} self-loop uncertainty $y_{sl}$ of input $x$.
    \end{algorithmic}
\end{algorithm}


The Jigsaw puzzles transformation adopted in our approach has two differences, compared to the one in \cite{noroozi2016jigsaw_puzzles}. First, to increase the diversity of permutation, each of the tiles is randomly rotated by an angle $a \in \{0^{\circ}, 90^{\circ}, 180^{\circ}, 270^{\circ}\}$ besides the translation transformation. Second, to integrate the Jigsaw puzzles task into the end-to-end semi-supervised framework, the input of self-supervised sub-task is required to have the same size as that of the target segmentation task. Hence, instead of using the shared-weight neural network for each tile, the permuted tiles are first assembled to an image of the same size of the original image (i.e., $\{P_1^{'},...,P_{Q}^{'}\}$ shown in Fig.~\ref{fig1:pipeline}) and then fed as input to the neural network for the permutation classification.

\subsection{Estimation of Self-loop Uncertainty for Unlabeled Data}
The generation procedure of our self-loop uncertainty is presented in Alg.~\ref{alg:ysl}. The self-supervised sub-task is able to recurrently optimize the neural network in an iteration, as the self-supervised signal can be self-driven without manual annotation. The different stages (i.e., $\{\theta_{e}^{i},\theta_{d}\}, i \in \{1,...,Q\}$) of self-supervised optimization are seen as different models, which enable the proposed self-loop uncertainty to approximate the ensemble uncertainty with a single neural network. The permutated images go through the FCN and yield a set of segmentation predictions $S_i, i \in \{1,...,Q\}$. Since the calculated self-supervised loss ($l$) can explicitly represent the difficulty of puzzled image for neural network to restore, we formulate $l$ as the confidence of corresponding segmentation result $S$ (via $norm(.)$ and $\omega$ defined in Alg.~\ref{alg:ysl}) to revise its contribution to the final pseudo-label. Our self-loop uncertainty thereby is the weighted average of the segmentation predictions produced by different stages of self-supervised optimization.

\paragraph{\bf Uncertainty-guided Loss.} The set of segmentation predictions $\{S_1,...,S_Q\}$ is presented in Fig.~\ref{fig1:pipeline}, where the red color represents the high score of foreground. The weight-averaged self-loop uncertainty $y_{sl}$ can be used as the guidance to maintain the reliable predictions (i.e., high score) as target for the neural network to learn from unlabeled data. To achieve this goal, we adopt the mean squared error (MSE) loss as the uncertainty-guided loss $\mathcal{L}_{UG}$ for the network optimization with unlabeled data and pseudo-labels $y_{sl}$, which can be defined as:

\begin{equation}
    \mathcal{L}_{UG}(S_x,y_{sl})=\frac{\sum_{H \times W}\mathbb{I}(y_{sl}>th)\| S_x - y_{sl}\|^{2}} {\sum_{H \times W} \mathbb{I}(y_{sl}>th)}
\end{equation}
where $\mathbb{I}$ is the indicator function; $H$ and $W$ are the image height and width, respectively; $S_x$ is the segmentation prediction of input image $x$; and $th$ is the threshold to select the high score target.

\subsection{Objective Function}
Assuming a batch contains $N$ labeled data ($\{(x_j,y_j)\}_{j=1}^N$) and $M$ unlabeled data $\{x_j\}_{j=N+1}^{N+M}$, where $x_{j} \in \mathbb{R}^{H \times W \times C}$ is the input image ($H$, $W$, and $C$ are the height, width, and channel of the image, respectively) and $y_{j} \in\{0,1\}^{H \times W}, j=1, 2, \dots, N$ is the ground-truth annotation, the objective function $\mathcal{L}$ for this batch can be formulated as:
\begin{equation}
    \mathcal{L} = \sum_{j=1}^{N} \mathcal{L}_{SEG}(x_j,y_j) + \sum_{j=N+1}^{N+M} \mathcal{L}_{UG}(x_j,y_{sl})+ \sum_{j=1}^{N+M} \sum_{i=1}^Q \mathcal{L}_{SS}(T_{P_i^{'}}(x_j), g_i).
\end{equation}

During network optimization, for the unlabeled data, we first fixed the decoder of FCN and recurrently update the encoder with $\mathcal{L}_{SS}$ to generate $y_{sl}$. Then, the weight of the whole FCN is optimized by $\mathcal{L}_{UG}$. In other words, an unsynchronized optimization of the encoder and decoder happens when using the unlabeled data. For the labeled data, on the other hand, the network is optimized with the $\mathcal{L}_{SEG}$ and $\mathcal{L}_{SS}$ simultaneously.


\section{Experiments}

\paragraph{\bf MoNuSeg Dataset \cite{naylor2018segmentation}.} The dataset consists of diverse H$\&$E stained tissue images captured from seven different organs (e.g., breast, liver, kidney, prostate, bladder, colon and stomach), which were collected from 18 institutes. The dataset has a public training set and a public test set for network training and evaluation, respectively. The training set contains 30 histopathological images with hand-annotated nuclei, while the test set consists of 14 images. The size of the histopathological images is $1000 \times 1000$ pixels.

\paragraph{\bf ISIC Dataset \cite{ISIC2019}.} The ISIC dataset is widely-used to assess the segmentation accuracy of skin lesion areas of automatic segmentation algorithms. The dataset contains 2,594 dermoscopic images. The skin lesion area of each image has been manually annotated by the data provider. The image size varies from around $1000 \times 1000$ pixels to $4000 \times 3000$ pixels. We resize all the images to a uniform size of $512 \times 512$ pixels for network training and validation. The dataset is randomly separated to training and test sets according to the ratio of 75:25.

\paragraph{\bf Evaluation Criterion.} The F1 score, i.e., the unweighted average classification accuracy of the foreground and background tissues, which is widely-used in the area of nuclei \cite{LunaM2019,OdaH2018,ZhouY2019} and skin lesion \cite{LiY2018,CMIG2019,TangY2019} segmentation, is adopted as the metric to evlauate the segmentation performance.

\paragraph{\bf Baselines.} Three popular uncertainty approaches---softmax probability map \cite{BaiWJ2017}, Monte Carlo (MC) dropout \cite{SedaiS2019,YuL2019}, and uncertainty estimation via ensembling networks \cite{LakshminarayananB2017}---are involved as baselines in this study. Similar to \cite{SedaiS2019}, we set the dropout rate to 0.2 and forward the image through the neural network for ten times to generate MC dropout uncertainty. The ensemble uncertainty is generated by ensembling ten models trained with different network initializations. Consistent with the baselines, we generate ten permutations for an image to iteratively optimize the neural network and accordingly yield the self-loop uncertainty. The widely used ResUNet-18 \cite{He01,Ronneberger01} is used as the backbone for uncertainty estimation. For fair comparison, all the baselines are trained according to the same protocol.


\subsection{Evaluation of Pseudo-label Quality}
Compared to skin lesion segmentation, which contains a single target in each image, the annotation of nucleus is more difficult and laborious. Hence, we mainly use the MoNuSeg dataset to evaluate the quality of pseudo-label yielded by different approaches in this section.\footnote{For visual comparison between pseudo-labels, please refer to {\itshape arxiv version.}}

To quantitatively validate the accuracy of different pseudo-labels, we calculate the F1 score between the pseudo-labels and ground-truth and present the results in Tabel~\ref{tab1:pseduolabelQ}. The pseudo-labels are generated with different amounts (i.e., 20\% and 50\%) of labeled data $D_L$ and the remaining training set is used as unlabeled data $D_U$. As shown in Table~\ref{tab1:pseduolabelQ}, our self-loop uncertainty outperforms all the baselines under different amounts of labeled data, which are $+2.27\%$ and $+2.88\%$ higher than the runner-up (i.e., MC Dropout) with 20\% and 50\% labeled data, respectively. The pesudo-labels yielded by uncertainty via ensembling models achieve lower accuracy among the baselines. The underlying reason may be that the MoNuSeg training set only contains 30 histopathological images, which make the amount of labeled data (i.e., 20\% and 50\%) insufficient to well train the neural network. Therefore, the ensembling of multiple unsatisfactory models cannot improve the accuracy of uncertainty estimation.

\begin{table}[!tb]
    \caption{F1 score (\%) between ground-truth and the pseudo-labels generated by different uncertainty approaches with different amounts of labeled data. The superscript of SL is the number of permutations $Q$ generated for self-supervised learning. (MC D.---MC Dropout, SL---Self-loop)}\label{tab1:pseduolabelQ}
    \begin{center}
        \small
        \begin{tabular}{c|cccccc}
            \hline
            {\bf Amount of $D_{L}$} & { \bf Softmax} & { \bf MC D.} & { \bf Ensemble} & {\,\,\, \bf SL$^{3}$ \,\,\,} & { \,\,\, \bf SL$^{6}$ \,\,\,} & { \,\,\, \bf SL$^{10}$ \,\,\,} \\
            \hline\hline
            {20\%}                  & 67.48          & 72.42        & 67.46           & 73.90           & 74.68           & \bf 75.24        \\
            \hline
            {50\%}                  & 69.53          & 73.58        & 70.01           & 76.51           & 76.77           & \bf 76.85        \\
            \hline
        \end{tabular}
    \end{center}
\end{table}

\paragraph{\bf Ablation Study.} We conduct an ablation study to investigate the relationship between the number of permutation $Q$ and the quality of pseudo-label. $Q$ is set to 3, 6, 10, respectively, for the generation of self-loop uncertainty. As shown in Table~\ref{tab1:pseduolabelQ}, the self-loop uncertainty generated with a larger $Q$ achieves the higher F1 score. However, the improvement of F1 score provided by increasing $Q$ from 6 to 10 becomes marginal (e.g., $+0.08\%$ using 50\% labeled data), which illustrates that $Q$ may not be the larger the better for practical applications, when taking the computational cost into account.

\subsection{Segmentation Performance Evaluation}
To validate the effectiveness of pseudo-labels, we evaluate the performance of different semi-supervised frameworks on the test sets of MoNuSeg and ISIC. The semi-supervied approaches are trained with different portions (i.e., 20\% and 50\%) of labeled data. The evaluation results are listed in Table~\ref{tab2:sota}. The performance of fully-supervised approach with 100\% labeled data is also assessed as the upper bound for the semi-supervised approaches. To validate the effectiveness of self-supervised sub-task, the self-loop uncertainty without $\mathcal{L}_{SS}$ is also involved for comparison. We pass the ten permutated images through the FCN without self-supervised optimization and yield the uncertainty by averaging the segmentation predictions. Due to lack of extra information exploited by self-supervised sub-task, the improvements yielded without $\mathcal{L}_{SS}$ significantly decrease.

\paragraph{\bf Nuclei Segmentation.} As shown in Table~\ref{tab2:sota}, the performance of fully-supervised approach significantly drops from 79.30\% to 75.87\% and 71.51\%, respectively, with the reductions (i.e., $-50\%$ and $-80\%$) of manual annotations. The application of pseudo-labels provides a consistent improvement to the segmentation accuracy. Among them, the proposed self-loop uncertainty yields the largest improvements, especially under the condition with 20\% annotated data, i.e., $+5.6\%$ higher than the fully-supervised approach. Furthermore, we notice that our semi-supervised framework trained with 50\% labeled data achieves comparable F1 score (79.10\%) to that of 100\% fully-supervised approach (79.30\%), which demonstrates the potential of our approach for reducing the workload of manual annotations.

\paragraph{\bf Skin Lesion Segmentation.} Similar trends of improvement are observed on the ISIC test set. Due to the extra information provided by the unlabeled data, the semi-supervised approaches outperform the fully-supervised one with limited annotated data (20\% and 50\%). The framework adopted our self-loop uncertainty as pseudo-labels achieves the highest F1 scores, i.e., 84.92\% and 86.17\% with 20\% and 50\% labeled data, respectively, and the latter is comparable to that of fully-supervied approach with 100\% annotations (i.e., 86.17\%). As ISIC has much more training data, compared to MoNuSeg, the ensemble-uncertainty-based framework achieves a comparable F1 score of 86.06\% with 50\% labeled data. However, it is worthwhile to mention that the generation of ensemble uncertainty requires 10 times of inferences during the test phase, as well as the MC dropout. Conversely, the proposed self-loop uncertainty can be generated with a single inference, which significantly saves the computational cost.

\begin{table}[!tb]
    \caption{F1 score (\%) yielded by different semi-supervised approaches on the two publicly available datasets.}\label{tab2:sota}
    \begin{center}
        \begin{tabular}{l|ccc|ccc}
            \hline
            {}                                            & \multicolumn{3}{c|}{\bf MoNuSeg} & \multicolumn{3}{c}{\bf ISIC}                                                                                                          \\\cline{2-7}
            {}                                            & {\,\,\,20\%\,\,\,}         & {\,\,\,50\%\,\,\,}           & {\,\,\,100\%\,\,\,} & {\,\,\,20\% \,\,\,} & {\,\,\,50\%\,\,\,} & {\,\,\,100\%\,\,\,} \\
            \hline\hline
            {\bf Fully-supervised}                        & 71.51                            & 75.87                        & \bf 79.30                     & 81.49                     & 84.86              & \bf 86.58                     \\\hline
            {{\bf Softmax} \cite{BaiWJ2017}}              & 73.65                            & 76.18                        & -                         & 82.81                     & 85.11              & -                         \\\hline
            {{\bf MC Dropout} \cite{SedaiS2019}\,\,\,}    & 75.31                            & 77.98                        & -                         & 83.68                     & 85.74              & -                         \\\hline
            {{\bf Ensemble} \cite{LakshminarayananB2017}} & 73.33                            & 76.87                        & -                         & 83.27                     & 86.06              & -                         \\\hline\hline
            {\bf Self-loop w/o $\mathcal{L}_{SS}$}        & 74.70                            & 77.78                        & -                         & 82.70                     & 85.22              & -                         \\\hline
            {\bf Self-loop (Ours)}                        & \bf 77.11                        & \bf 79.10                    & -                         & \bf 84.92                 & \bf 86.17          & -                         \\\hline
        \end{tabular}
    \end{center}
\end{table}

\section{Conclusion}
In this paper, we proposed a semi-supervised approach to train neural networks with limited labeled data and a large quantity of unlabeled images for medical image segmentation. A novel pseudo-label (namely self-loop uncertainty), generated by recurrently optimizing the neural network with a self-supervised task, is adopted as the ground-truth for the unlabeled images to augment the training set and boost the segmentation accuracy. 

\section*{Acknowledge}
This work is supported by the Natural Science Foundation of China (No. 61702339), the Key Area Research and Development Program of Guangdong Province, China (No. 2018B010111001), National Key Research and Development Project (2018YFC2000702) and Science and Technology Program of Shenzhen, China (No. ZDSYS201802021814180).


%
%
%
\bibliographystyle{splncs04}
\bibliography{my_reference}

\begin{thebibliography}{10}
\providecommand{\url}[1]{\texttt{#1}}
\providecommand{\urlprefix}{URL }
\providecommand{\doi}[1]{https://doi.org/#1}

\bibitem{BaiWJ2017}
Bai, W., Oktay, O., Sinclair, M., Suzuki, H., Rajchl, M., Tarroni, G., Glocker,
  B., King, A., Matthews, P.M., Rueckert, D.: Semi-supervised learning for
  network-based cardiac {MR} image segmentation. In: {International Conference
  on Medical Image Computing \& Computer Assisted Intervention}. pp. 253--260
  (2017)

\bibitem{BortsovaG2019}
Bortsova, G., Dubost, F., Hogeweg, L., Katramados, I., de~Bruijne, M.:
  Semi-supervised medical image segmentation via learning consistency under
  transformations. In: {International Conference on Medical Image Computing \&
  Computer Assisted Intervention}. pp. 810--818 (2019)

\bibitem{ISIC2019}
Codella, N., Rotemberg, V., Tschandl, P., Celebi, M.E., Dusza, S., Gutman, D.,
  Helba, B., Kalloo, A., Liopyris, K., Marchetti, M., Kittler, H., Halpern, A.:
  Skin lesion analysis toward melanoma detection 2018: {A} challenge hosted by
  the {I}nternational {S}kin {I}maging {C}ollaboration ({ISIC}). arXiv preprint
  arXiv:1902.03368  (2019)

\bibitem{GalY2016}
Gal, Y., Ghahramani, Z.: Dropout as a {B}ayesian approximation: {R}epresenting
  model uncertainty in deep learning. In: {International Conference on Machine
  Learning}. pp. 1050--1059 (2016)

\bibitem{gidaris2018image_rotations}
Gidaris, S., Singh, P., Komodakis, N.: Unsupervised representation learning by
  predicting image rotations. In: International Conference on Learning
  Representations (2018)

\bibitem{He01}
He, K., Zhang, X., Ren, S., Sun, J.: Deep residual learning for image
  recognition. In: IEEE Conference on Computer Vision and Pattern Recognition.
  pp. 770--778 (2016)

\bibitem{JungoA2019}
Jungo, A., Reyes, M.: Assessing reliability and challenges of uncertainty
  estimations for medical image segmentation. In: {International Conference on
  Medical Image Computing \& Computer Assisted Intervention}. pp. 48--56 (2019)

\bibitem{LakshminarayananB2017}
Lakshminarayanan, B., Pritzel, A., Blundell, C.: Simple and scalable predictive
  uncertainty estimation using deep ensembles. In: {Annual Conference on Neural
  Information Processing Systems}. pp. 6402--6413 (2017)

\bibitem{larsson_colorization_2017}
Larsson, G., Maire, M., Shakhnarovich, G.: Colorization as a proxy task for
  visual understanding. In: IEEE Conference on Computer Vision and Pattern
  Recognition. pp. 840--849 (2017)

\bibitem{LiY2018}
Li, Y., Shen, L.: Skin lesion analysis towards melanoma detection using deep
  learning network. Sensors  \textbf{18}(2), ~556 (2018)

\bibitem{LunaM2019}
Luna, M., Kwon, M., Park, S.H.: Precise separation of adjacent nuclei using a
  {S}iamese neural network. In: International Conference on Medical Image
  Computing \& Computer Assisted Intervention. pp. 577--585 (2019)

\bibitem{CMIG2019}
Nasr-Esfahani, E., Rafiei, S., Jafari, M.H., Karimi, N., Wrobel, J.S., Samavi,
  S., Soroushmehr, S.R.: Dense pooling layers in fully convolutional network
  for skin lesion segmentation. Computerized Medical Imaging and Graphics
  \textbf{78},  101658 (2019)

\bibitem{naylor2018segmentation}
Naylor, P., Lae, M., Reyal, F., Walter, T.: Segmentation of nuclei in
  histopathology images by deep regression of the distance map. IEEE
  Transactions on Medical Imaging  \textbf{38}(2),  448--459 (2018)

\bibitem{noroozi2016jigsaw_puzzles}
Noroozi, M., Favaro, P.: Unsupervised learning of visual representations by
  solving {J}igsaw puzzles. In: European Conference on Computer Vision. pp.
  69--84 (2016)

\bibitem{OdaH2018}
Oda, H., Roth, H., Chiba, K., Sokolic, J., Kitasaka, T., Oda, M., Hinoki, A.,
  Uchida, H., Schnabel, J., Mori, K.: {BESNet}: {B}oundary-enhanced
  segmentation of cells in histopathological images. In: International
  Conference on Medical Image Computing \& Computer Assisted Intervention. pp.
  228--236 (2018)

\bibitem{Ronneberger01}
Ronneberger, O., Fischer, P., Brox, T.: {U-Net}: Convolutional networks for
  biomedical image segmentation. In: International Conference on Medical Image
  Computing \& Computer Assisted Intervention. pp. 234--241 (2015)

\bibitem{SedaiS2019}
Sedai, S., Antony, B., Rai, R., Jones, K., Ishikawa, H., Schuman, J., Gadi, W.,
  Garnavi, R.: Uncertainty guided semi-supervised segmentation of retinal
  layers in {OCT} images. In: {International Conference on Medical Image
  Computing \& Computer Assisted Intervention}. pp. 282--290 (2019)

\bibitem{TangY2019}
{Tang}, Y., {Yang}, F., {Yuan}, S., {Zhan}, C.: A multi-stage framework with
  context information fusion structure for skin lesion segmentation. In:
  International Symposium on Biomedical Imaging. pp. 1407--1410 (2019)

\bibitem{YuL2019}
Yu, L., Wang, S., Li, X., Fu, C.W., Heng, P.A.: Uncertainty-aware
  self-ensembling model for semi-supervised {3D} left atrium segmentation. In:
  {International Conference on Medical Image Computing \& Computer Assisted
  Intervention}. pp. 605--613 (2019)

\bibitem{ZhouY2019}
Zhou, Y., Onder, O.F., Dou, Q., Tsougenis, E., Chen, H., Heng, P.A.: {CIA-Net}:
  {R}obust nuclei instance segmentation with contour-aware information
  aggregation. In: {International Conference on Information Processing in
  Medical Imaging}. pp. 682--693 (2019)

\end{thebibliography}

\clearpage
\section*{Appendix}


\begin{figure}[!htb]
    \includegraphics[width=\textwidth]{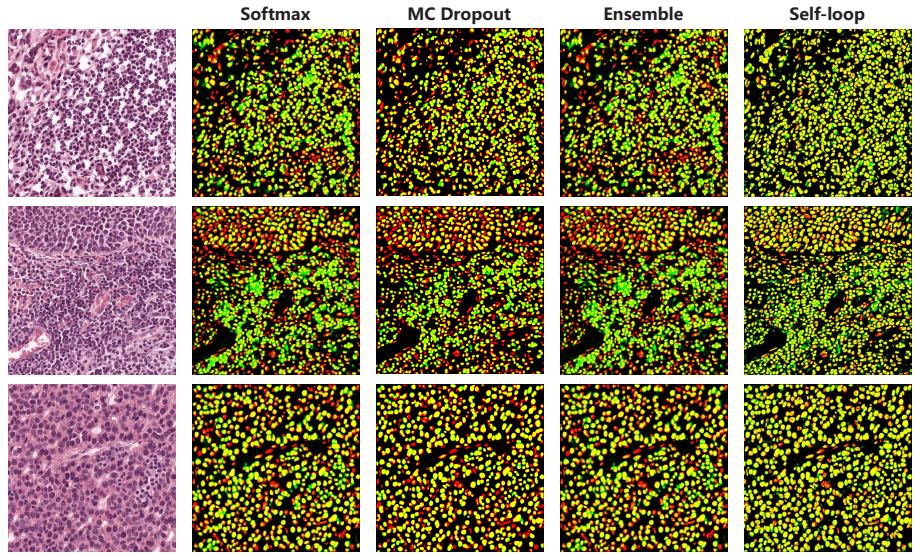}
    \caption{Visualization of pseudo-label yielded by different uncertainty approaches with {\bf 20\% labeled data}. The pseudo-labels and ground-truth are in \textcolor{green}{green} and \textcolor{red}{red}, respectively. The overlapping areas are in \textcolor{yellow}{yellow}. It can be observed that our self-loop uncertainty is closer to the ground-truth (i.e., larger overlapping areas), compared to other approaches.} \label{fig2:visualization}
\end{figure}

\begin{figure}[!htb]
    \includegraphics[width=\textwidth]{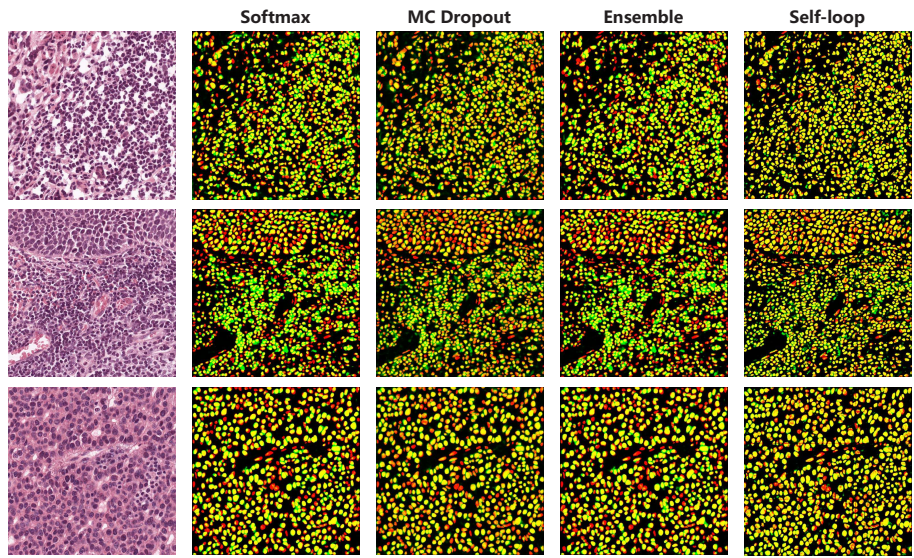}
    \caption{Visualization of pseudo-label yielded by different uncertainty approaches with {\bf 50\% labeled data}. The pseudo-labels and ground-truth are in \textcolor{green}{green} and \textcolor{red}{red}, respectively. The overlapping areas are in \textcolor{yellow}{yellow}. The proposed self-loop uncertainty achieves larger overlapping areas than the others.} \label{fig2:visualization_2}
\end{figure}

\end{document}